\documentclass[aps,prl,reprint,amsmath,amssymb,superscriptaddress,showpacs,twocolumn,nofootinbib,nobibnotes,nobibnotes]{revtex4-2}

\usepackage{graphicx}
\usepackage{dcolumn}
\usepackage{multirow}
\usepackage{bm}
\usepackage{tensor}
\usepackage{textcomp,mathcomp}
\usepackage{float}
\usepackage{color}
\usepackage{threeparttable}
\usepackage{xcolor} 
\definecolor{darkblue}{rgb}{0.176,0.184,0.569} 
\usepackage[colorlinks,allcolors=darkblue]{hyperref}
\usepackage{CJK}
\usepackage{url}
\usepackage{supertabular}
\usepackage{color}
\usepackage{makecell}
\usepackage{lipsum}

\begin{document}


\title{Enhanced $S$-factor for the $^{14}$N$(p,\gamma)^{15}$O reaction and its impact on the solar composition problem}


\author{X.~Chen}
\affiliation{Key Laboratory of Beam Technology of Ministry of Education, \\School of Physics and Astronomy, Beijing Normal University, Beijing 100875, China}

\author{J.~Su}
\email[]{sujun@bnu.edu.cn}
\affiliation{Key Laboratory of Beam Technology of Ministry of Education, \\School of Physics and Astronomy, Beijing Normal University, Beijing 100875, China}

\author{Y.~P.~Shen}
\email[]{ypshen@ciae.ac.cn}
\affiliation{China Institute of Atomic Energy, P. O. Box 275(10), Beijing 102413, China}

\author{L.~Y.~Zhang}
\affiliation{Key Laboratory of Beam Technology of Ministry of Education, \\School of Physics and Astronomy, Beijing Normal University, Beijing 100875, China}

\author{J.~J.~He}
\affiliation{Key Laboratory of Nuclear Physics and Ion-beam Application (MoE), Institute of Modern Physics, Fudan University, Shanghai, 200433, China}
\affiliation{Key Laboratory of Beam Technology of Ministry of Education, \\School of Physics and Astronomy, Beijing Normal University, Beijing 100875, China}

\author{S.~Z.~Chen}
\affiliation{Institute of Nuclear Energy Safety Technology of Hefei Institutes of Physical Science, Chinese Academy of Science, Hefei 230031, China}

\author{S.~Wang~}
\affiliation{Shandong Provincial Key Laboratory of Optical Astronomy and Solar-Terrestrial Environment, Institute of Space Sciences, Shandong University, Weihai 264209, China}

\author{Z.~L.~Shen}
\affiliation{Key Laboratory of Beam Technology of Ministry of Education, \\School of Physics and Astronomy, Beijing Normal University, Beijing 100875, China}

\author{S.~Lin}
\affiliation{Key Laboratory of Beam Technology of Ministry of Education, \\School of Physics and Astronomy, Beijing Normal University, Beijing 100875, China}

\author{L.~Y.~Song}
\affiliation{Key Laboratory of Beam Technology of Ministry of Education, \\School of Physics and Astronomy, Beijing Normal University, Beijing 100875, China}

\author{H.~Zhang}
\affiliation{Key Laboratory of Nuclear Physics and Ion-beam Application (MoE), Institute of Modern Physics, Fudan University, Shanghai, 200433, China}

\author{L.~H.~Wang}
\affiliation{School of Mathematics and Physics, Handan University, Handan 056005, China}

\author{X.~Z.~Jiang}
\affiliation{Key Laboratory of Beam Technology of Ministry of Education, \\School of Physics and Astronomy, Beijing Normal University, Beijing 100875, China}

\author{L.~Wang}
\affiliation{Key Laboratory of Beam Technology of Ministry of Education, \\School of Physics and Astronomy, Beijing Normal University, Beijing 100875, China}

\author{Y.~T.~Huang}
\affiliation{Key Laboratory of Beam Technology of Ministry of Education, \\School of Physics and Astronomy, Beijing Normal University, Beijing 100875, China}

\author{Z.~W.~Qin}
\affiliation{Key Laboratory of Beam Technology of Ministry of Education, \\School of Physics and Astronomy, Beijing Normal University, Beijing 100875, China}

\author{F.~C.~Liu}
\affiliation{Key Laboratory of Beam Technology of Ministry of Education, \\School of Physics and Astronomy, Beijing Normal University, Beijing 100875, China}

\author{Y.~D.~Sheng}
\affiliation{Key Laboratory of Beam Technology of Ministry of Education, \\School of Physics and Astronomy, Beijing Normal University, Beijing 100875, China}

\author{Y.~J.~Chen}
\affiliation{Key Laboratory of Nuclear Physics and Ion-beam Application (MoE), Institute of Modern Physics, Fudan University, Shanghai, 200433, China}

\author{Y.~L.~Lu}
\affiliation{Key Laboratory of Nuclear Physics and Ion-beam Application (MoE), Institute of Modern Physics, Fudan University, Shanghai, 200433, China}

\author{X.~Y.~Li}
\affiliation{Key Laboratory of Nuclear Physics and Ion-beam Application (MoE), Institute of Modern Physics, Fudan University, Shanghai, 200433, China}

\author{J.~Y.~Dong}
\affiliation{China Institute of Atomic Energy, P. O. Box 275(10), Beijing 102413, China}

\author{Y.~C.~Jiang}
\affiliation{China Institute of Atomic Energy, P. O. Box 275(10), Beijing 102413, China}

\author{Y.~Q.~Zhang}
\affiliation{China Institute of Atomic Energy, P. O. Box 275(10), Beijing 102413, China}

\author{Y.~Zhang}
\affiliation{China Institute of Atomic Energy, P. O. Box 275(10), Beijing 102413, China}

\author{J.~W.~Tian}
\affiliation{China Institute of Atomic Energy, P. O. Box 275(10), Beijing 102413, China}

\author{D.~Xiao}
\affiliation{Institute of Nuclear Energy Safety Technology of Hefei Institutes of Physical Science, Chinese Academy of Science, Hefei 230031, China}

\author{Y.~Zhang}
\affiliation{Institute of Nuclear Energy Safety Technology of Hefei Institutes of Physical Science, Chinese Academy of Science, Hefei 230031, China}

\author{Z.~M.~Li}
\affiliation{Harbin Engineering University, Harbin, Heilongjiang, 150000, China}

\author{X.~C.~Han}
\affiliation{Shandong Provincial Key Laboratory of Optical Astronomy and Solar-Terrestrial Environment, Institute of 
Space Sciences, Shandong University, Weihai 264209, China}

\author{J.~J.~Wei}
\affiliation{Institute for Advanced Materials and Technology, University of Science and Technology Beijing,  Beijing 100083, China}

\author{H.~Li}
\affiliation{Institute for Advanced Materials and Technology, University of Science and Technology Beijing,  Beijing 100083, China}

\author{Z.~An}
\affiliation{Key Laboratory of Radiation Physics and Technology of the Ministry of Education, Institute of Nuclear Science and Technology,
Sichuan University, Chengdu 610064, China}

\author{W.~P.~Lin}
\affiliation{Key Laboratory of Radiation Physics and Technology of the Ministry of Education, Institute of Nuclear Science and Technology,
Sichuan University, Chengdu 610064, China}

\author{B.~Liao}
\affiliation{Key Laboratory of Beam Technology of Ministry of Education, \\School of Physics and Astronomy, Beijing Normal University, Beijing 100875, China}

\author{H.~N.~Liu}
\affiliation{Key Laboratory of Beam Technology of Ministry of Education, \\School of Physics and Astronomy, Beijing Normal University, Beijing 100875, China}

\author{F.~S.~Zhang}
\affiliation{Key Laboratory of Beam Technology of Ministry of Education, \\School of Physics and Astronomy, Beijing Normal University, Beijing 100875, China}

\author{M.~L.~Qiu}
\affiliation{Key Laboratory of Beam Technology of Ministry of Education, \\School of Physics and Astronomy, Beijing Normal University, Beijing 100875, China}

\author{C.~Xu}
\affiliation{Key Laboratory of Beam Technology of Ministry of Education, \\School of Physics and Astronomy, Beijing Normal University, Beijing 100875, China}

\author{S.~L.~Jin}
\affiliation{Institute of Modern Physics, Chinese Academy of Sciences, Lanzhou, 730000, China}

\author{F.~Lu}
\affiliation{Shanghai Advanced Research Institute, Chinese Academy of Sciences, Shanghai 201210, China}

\author{J.~F.~Chen}
\affiliation{Shanghai Institute of Ceramics, Chinese Academy of Sciences, Shanghai 201800, China}

\author{W.~Nan}
\affiliation{China Institute of Atomic Energy, P. O. Box 275(10), Beijing 102413, China}

\author{Y.~B.~Wang}
\affiliation{China Institute of Atomic Energy, P. O. Box 275(10), Beijing 102413, China}

\author{Z.~H.~Li}
\affiliation{China Institute of Atomic Energy, P. O. Box 275(10), Beijing 102413, China}

\author{B.~Guo}
\affiliation{China Institute of Atomic Energy, P. O. Box 275(10), Beijing 102413, China}

\author{Y.~Q.~Gu}
\affiliation{Science and Technology on Plasma Physics Laboratory, Laser Fusion Research Center,
China Academy of Engineering Physics, Mianyang 621900, China}

\author{W.~P.~Liu}
\email[]{liuwp@sustech.edu.cn}
\affiliation{Department of Physics, Southern University of Science and Technology, Shenzhen 518055, China}
\affiliation{China Institute of Atomic Energy, P. O. Box 275(10), Beijing 102413, China}


\date{\today}

\begin{abstract}
The solar composition problem has puzzled astrophysicists for more than 20 years. Recent measurements of carbon-nitrogen-oxygen (CNO) neutrinos by the Borexino experiment show a $\sim2\sigma$ tension with the ``low-metallicity'' determinations. $^{14}$N$(p,\gamma)^{15}$O, the slowest reaction in the CNO cycle, plays a crucial role in the standard solar model (SSM) calculations of CNO neutrino fluxes. Here we report a direct measurement of the $^{14}$N$(p,\gamma)^{15}$O reaction, in which $S$-factors for all transitions were simultaneously determined in the energy range of $E_p=110-260$~keV for the first time. Our results resolve previous discrepancies in the ground-state transition, yielding a zero-energy $S$-factor $S_{114}(0) = 1.92\pm0.08$~keV~b which is 14\% higher than the $1.68\pm0.14$~keV~b recommended in Solar Fusion III (SF-III). With our $S_{114}$ values, the SSM B23-GS98, and the latest global analysis of solar neutrino measurements, the C and N photospheric abundance determined by the Borexino experiment is updated to $N_{\mathrm{CN}}=({4.45}^{+0.69}_{-0.61})\times10^{-4}$. This new $N_{\mathrm{CN}}$ value agrees well with latest ``high-metallicity'' composition, however, is also consistent with the ``low-metallicity'' determination within $\sim 1\sigma$ C.L., indicating that the solar metallicity problem remains an open question. In addition, the significant reduction in the uncertainty of $S_{114}$ paves the way for the precise determination of the CN abundance in future large-volume solar neutrino measurements.
\end{abstract}

\maketitle



The study of solar properties not only provides a benchmark for exploring the vast stellar domains of the cosmos, but also advances a broad range of related disciplines. Solar metallicity, defined as the ratio of elements heavier than helium to hydrogen in the Sun, serves as a key parameter in the standard solar models (SSMs), directly influencing our understanding of solar properties. Early solar metallicities determined by modeling the spectra observed from the solar photosphere agreed with estimates from helioseimology measurements. However, the metallicities reported by the works~\cite{Asplund2009,caffau2011solar,asplund2021chemical} using the advanced techniques consistently exhibit a 30\% to 40\% reduction compared to the value given by traditional modeling~\cite{Grevesse1998}. This discrepancy, known as the ``solar composition problem'' or ``solar abundance problem'', has puzzled astrophysicists for over two decades, prompting ongoing investigations from multiple perspectives. In this letter, we follow the statements in Solar Fusion III (SF-III)~\cite{acharya2024solar} and adopt the metallicities from \citet{asplund2021chemical} and \citet{magg2022observational} to represent the low-metallicity (LZ) and high-metallicity (HZ) compositions, respectively.

Solar neutrino measurements provide direct insights into the solar interior, opening a new approach to resolving the solar composition problem. Among the various types of solar neutrinos, carbon-nitrogen-oxygen (CNO) neutrinos are considered the most effective tool for studying solar metallicity~\cite{haxton2008cn,Serenelli2013}. Remarkable progress has been achieved by recent Borexino experiments~\cite{Borexino2020, Borexino2022}. The measured CNO neutrino flux, combined with the predictions of SSMs, has been used to derive the CN abundance with respect to the H abundance $N_{\rm{CN}}$ in the photosphere of the sun. The result, $N_{\rm{CN}}=(5.81^{+1.22}_{-0.94}\times10^{-4})$~\cite{Basilico23}, shows a preference for HZ compositions and a $\sim2\sigma$ tension with the LZ determinations.  

$^{14}$N$(p,\gamma)^{15}$O, the slowest reaction in the CNO cycle, directly controls the production rate of CNO neutrinos~\cite{Wiescher2018}. Consequently, the $S$-factor of this reaction, $S_{114}$, plays a crucial role in the SSMs calculation and the estimation of the $N_{\rm{CN}}$. Furthermore, due to the crucial influence on the timescale of hydrogen burning in stars more massive than the Sun~\cite{Wiescher2010}, precise $S_{114}$ values can help us better understand the age and evolution of star clusters and galaxies~\cite{Chaboyer_1998,Science1996}. After extensive efforts over the past decades, $S_{114}$ values have been significantly improved. The latest compilation SF-III~\cite{acharya2024solar} recommended a zero energy $S$-factor of $S_{114}(0)=1.68\pm0.14$~keV~b. Although it represents a secondary contribution to the uncertainty in estimating $N_{\rm{CN}}$, this 8.4\% error will become a bottleneck as CN neutrino data are improved by future large-volume experiments~\cite{acharya2024solar}.

Currently, the improvement in the precision of $S_{114}$ is limited by the incomplete understanding of the ground-state transition at low energies, which are strongly influenced by a sub-threshold resonance at $E_x=6.79$~MeV. In 1987, \citet{schroder1987stellar} obtained a large zero-energy $S$-factor for the ground-state transition $S_{114}^{\rm{g.s.}}(0)= 1.55\pm0.34$~keV~b from direct measurement. However, this result was dramatically revised to $0.08^{+0.13}_{-0.06}$~keV~b by \citet{angulo200114n}. In 2005, two direct measurements were performed by \citet{Runkle2005} and \citet{imbriani2005s} at lower energies, respectively. The $S_{114}^{\rm{g.s.}}(0)= 0.49\pm0.08$~keV~b obtained in \citet{Runkle2005} is significantly higher than the $S_{114}^{\rm{g.s.}}(0)= 0.25\pm0.06$~keV~b reported in \citet{imbriani2005s}. The main discrepancy between these two measurements~\cite{Runkle2005,imbriani2005s} lies in the trend of the $S$-factor around $E_\mathrm{c.m.}=200$ keV.
Several measurements~\cite{Marta2008,Li2016,Wagner2018,Frentz2022} of excitation function and angular distributions at higher energies have provided additional constraints for $R$-matrix fitting. Moreover, efforts have been made to measure the lifetime and width of the 6.79 MeV state in $^{15}$O~\cite{Berton2001,YAMADA2004265,Schurmann2008,Galinski2014,Sharma2020,Frentz2021}, but these efforts have not yet led to significant progress. 

In this Letter, we report a direct measurement of the $^{14}$N$(p,\gamma)^{15}$O cross sections at low energies. The $S$-factors for all transitions were simultaneously determined through a multi-channel Bayesian analysis of the obtained $\gamma$ spectra. Our results resolve the discrepancies in the ground-state transition and provide better constraints for $R$-matrix fitting. The updated $S_{114}$ values were then used to investigate the solar composition problem.

The experiment was conducted at Hefei 350~kV accelerator facility~\cite{CHEN2023168401} with a $\sim2$~mA proton beam and TiN targets. The $\gamma$ spectra were measured at $E_p=110-260$~keV using the second configuration of the Large Modular BGO Detector Array (LAMBDA-II) with a summing method~\cite{su2022, Wang23, ZH-25Mg}. The experimental setup and detailed procedures are provided in the Supplemental Material.

\begin{figure}[htbp]
\centering
\includegraphics[width=1.0\columnwidth]{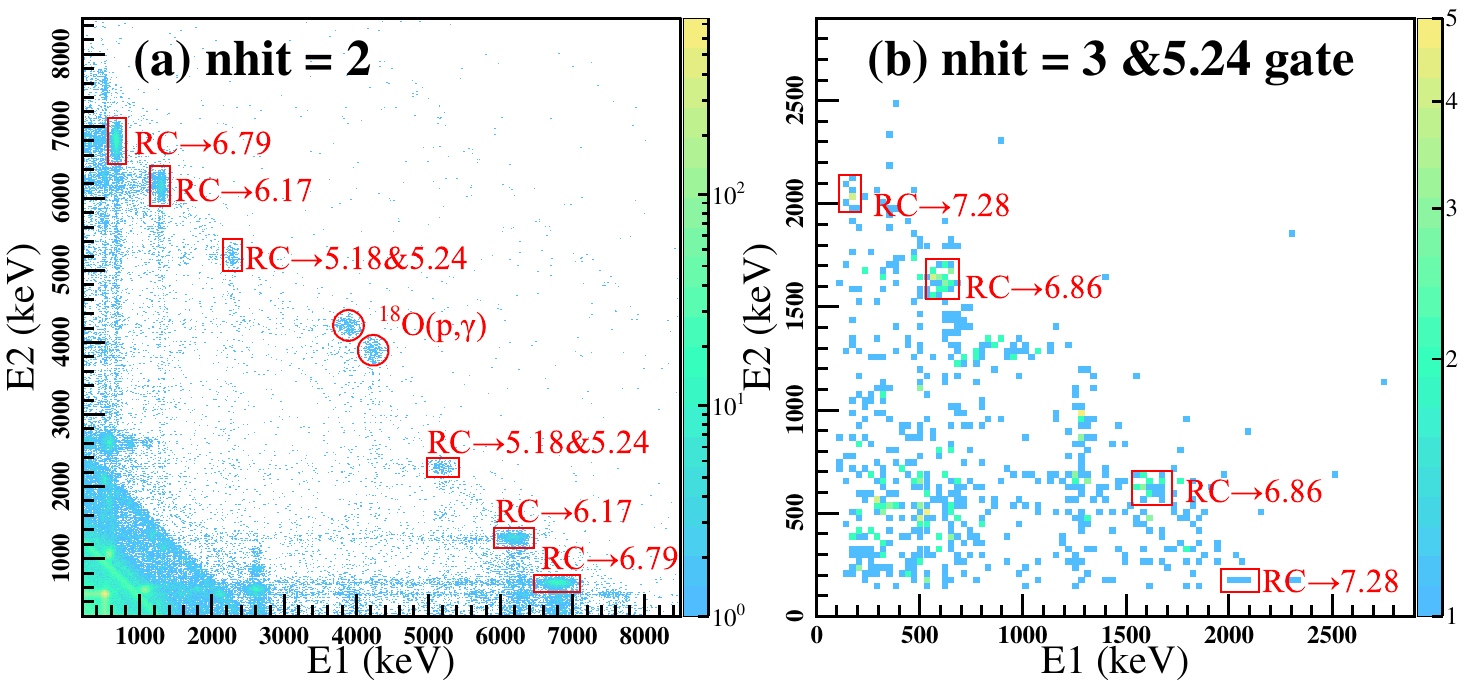}
\caption{$\gamma-\gamma$ coincidence spectra obtained at $E_p=180$~keV. (a) $\gamma-\gamma$ spectrum gated with nhit~=~2. (b) $\gamma-\gamma$ spectrum gated with nhit~=~3 and the third $\gamma$-ray falls into the $5.24\pm0.25$~MeV window. The coincidence windows of cascade $\gamma$ transitions of $^{14}$N$(p,\gamma)^{15}$O are marked with red boxes.}
\label{fig:gg} 
\end{figure}

In the data analysis, improvements were made to the previously used summing method~\cite{su2022, Wang23, ZH-25Mg}. By taking advantage of the high granularity of LAMBDA-II (16 modules), the $\gamma-\gamma$ coincidences and number of triggered modules (nhit) were used to distinguish the events from different transitions in $^{14}$N$(p,\gamma)^{15}$O. These constraints helped further reduce the interference from laboratory background, beam-induced background and other transitions. Figure \ref{fig:gg} shows the $\gamma-\gamma$ coincidence spectra obtained at $E_p=180$~keV. Four transitions following double cascades, including RC$\rightarrow$6.79$\rightarrow$0, RC$\rightarrow$6.17$\rightarrow$0, RC$\rightarrow$5.24$\rightarrow$0 and RC$\rightarrow$5.18$\rightarrow$0 can be clearly identified in Fig.~\ref{fig:gg}(a) gated by nhit=2. Two transitions following triple cascades, RC$\rightarrow$7.28$\rightarrow$5.24$\rightarrow$0 and RC$\rightarrow$6.86$\rightarrow$5.24$\rightarrow$0, are also identified in the $\gamma-\gamma$ coincidence spectrum gated by nhit=3, with an additional $5.24\pm0.25$~MeV gate on one of the triggered modules, as shown in Fig.~\ref{fig:gg}(b).  

\begin{figure*}[htpb]
\centering
\includegraphics[width=1.0\textwidth]{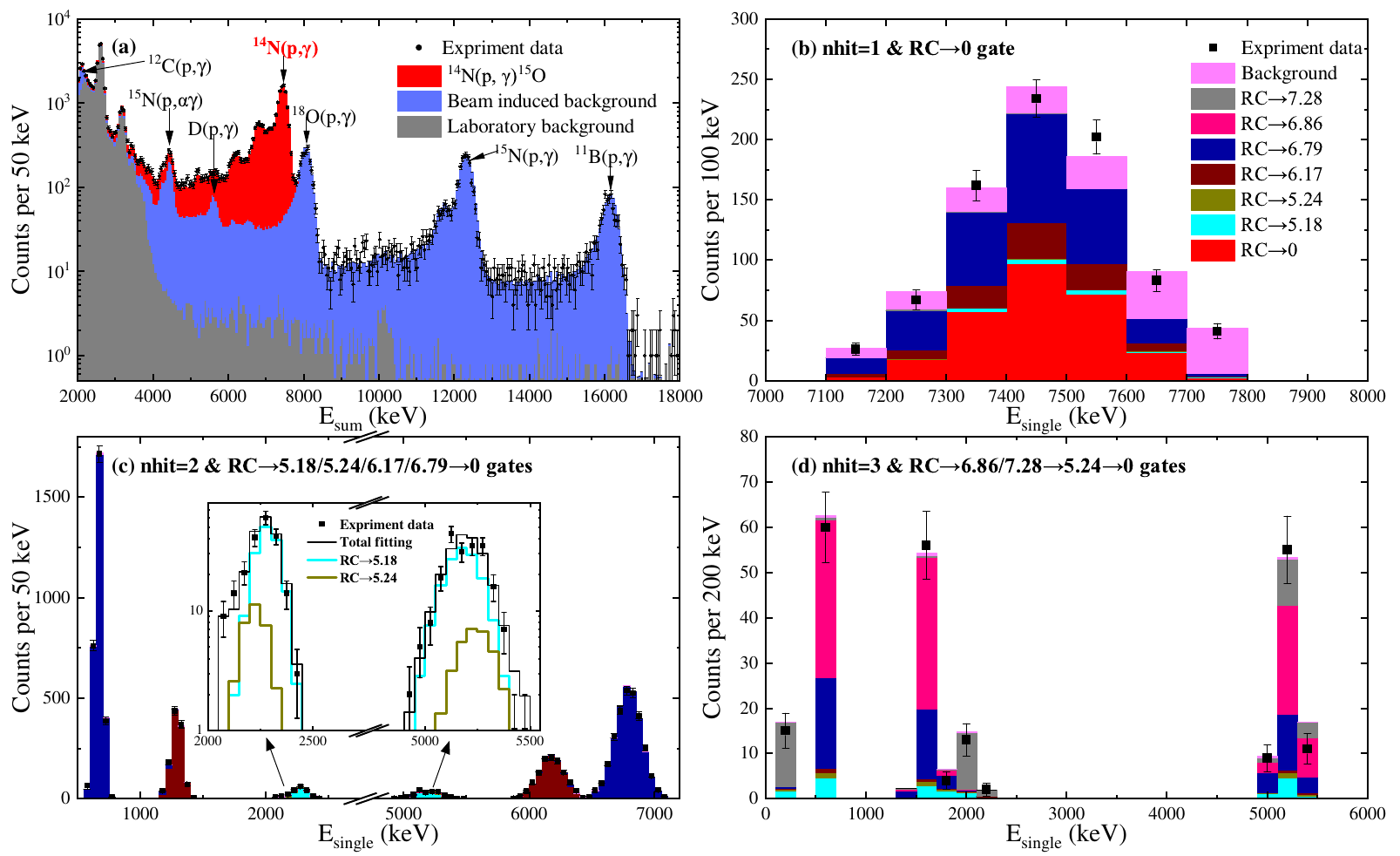}
\caption{Bayesian analysis of the spectra obtained at $E_p=180$~keV. (a) Sum spectrum. (b) Single spectrum gated by nhit~=~1 and the $\gamma$ peak of RC$\rightarrow$0. (c) Single spectrum gated by nhit~=~2 and $\gamma$-$\gamma$ coincidence gates for RC$\rightarrow$5.18/5.24/6.17/6.79 shown in Fig.~\ref{fig:gg}(a). In the inset the shapes of the 2.3 and 5.2~MeV peaks are expanded. (d) Single spectrum gated by nhit~=~3 and $\gamma$-$\gamma$ coincidence gates for RC$\rightarrow$6.86/7.28 shown in Fig.~\ref{fig:gg}(b). The different colored areas represent fractional contributions from the background and transitions of $^{14}$N$(p,\gamma)^{15}$O. For clarity, the contributions of each component were combined differently in the sum and single spectra.}
\label{fig:nhit-spectra} 
\end{figure*}

Figure \ref{fig:nhit-spectra} shows the sum spectrum (the sum of the energies) and the single spectrum (superimposing the spectra from each module) spectra obtained at $E_p=180$~keV. The single spectra were gated by different nhit values and corresponding $\gamma-\gamma$ coincidence gates. 

We applied a Bayesian approach to perform a multichannel fit to the four $\gamma$ spectra shown in Fig.~\ref{fig:nhit-spectra}, using the Bayesian Analysis Toolkit (BAT)~\cite{caldwell2010}. The shapes of fractional contribution from each transition to the corresponding spectrum were simulated using the GEANT4 program~\cite{ago03,all06,all16}. Both the laboratory background and beam-induced background were included in the fitting, as shown in Fig.~\ref{fig:nhit-spectra}(a). The laboratory background was reduced to a negligible level ($\sim1\times10^{-2}$~keV$^{-1}$h$^{-1}$ around the sum peak of $^{14}$N$(p,\gamma)^{15}$O) by several measures detailed in the Supplemental Material, except for the measurements at the lowest few energy points. The beam-induced background was obtained using corresponding targets at the same beam energy, with the exception of D$(p,\gamma)^{3}$He, which was constructed via Monte Carlo simulation.  

The fractional contributions from different transitions and background fitted using BAT are represented by different colored areas in Fig.~\ref{fig:nhit-spectra}. The single peaks corresponding to the 5.18, 5.24, 6.17, and 6.79~MeV transitions are clearly visible in Fig.~\ref{fig:nhit-spectra}(c), with minimal interference from other transitions. Although the contributions of the 5.18 and 5.24 transitions cannot be directly separated, they were effectively constrained by the shapes of the $\gamma$ peaks around 2.3 and 5.2~MeV, as shown in the inset of Fig.~\ref{fig:nhit-spectra}(c). Fig.~\ref{fig:nhit-spectra}(d) presents the fitting of the 6.86 and 7.28~MeV transitions following triple cascades, where considerable disturbances from the 6.79 and 5.18 transitions are found. Fig.~\ref{fig:nhit-spectra}(b) illustrates the fitting of the single peak corresponding to the ground-state transition, with approximately half of the contribution arises from the sum-in effect of other transitions. To validate the accuracy of the sum-in effect estimation, the branching ratios for the $E_p=278$~keV resonance were analyzed with the same process. As listed in the Supplemental Material, the obtained branching ratios are in agreement with those reported in previous works~\cite{Runkle2005,imbriani2005s,Marta2008,Daigle16}, including the ground-state transition. 

\begin{figure}[htpb]
\centering
\includegraphics[width=\columnwidth]{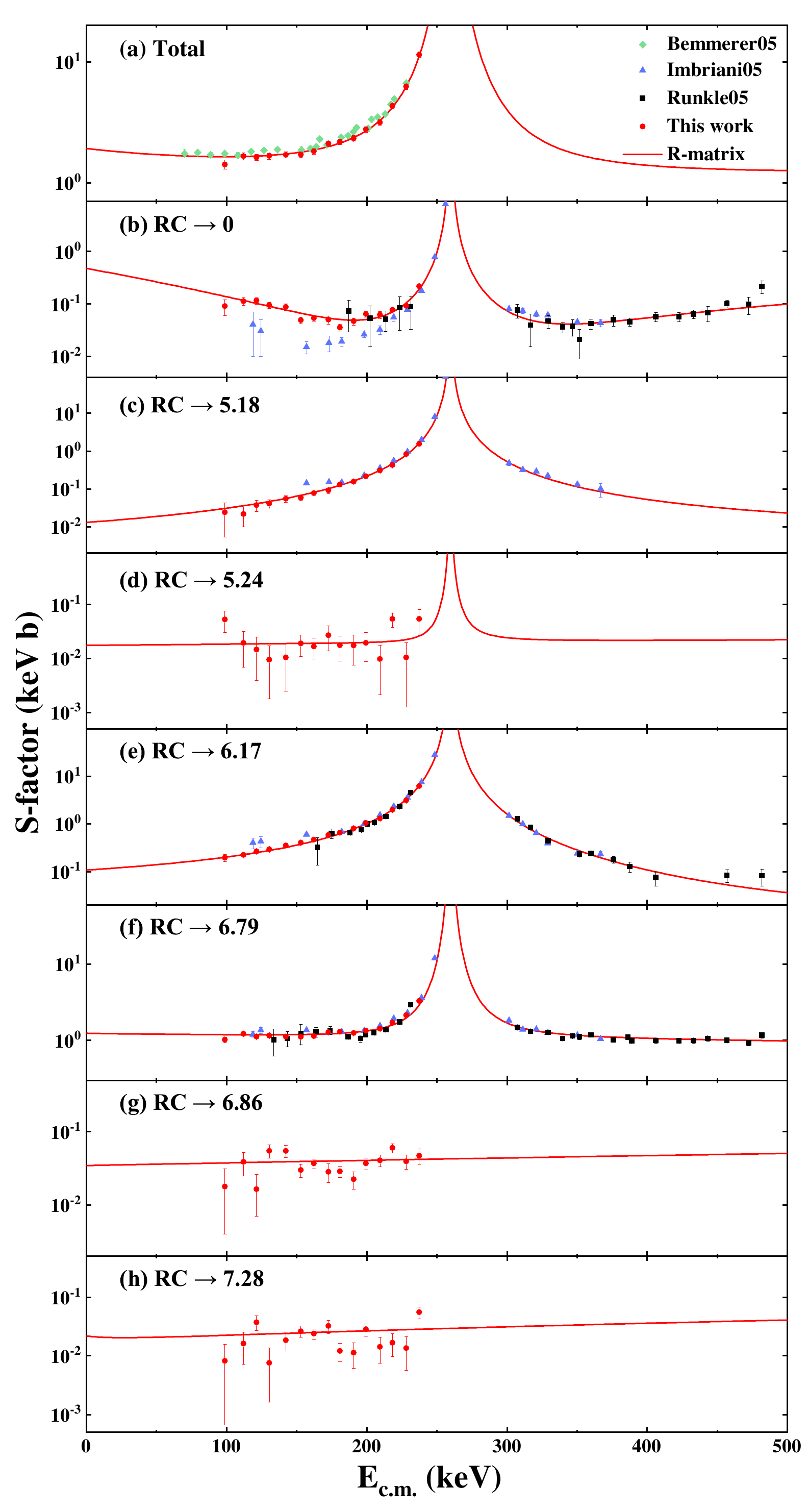}
\caption{(a) The total $S$-factors of $^{14}$N$(p,\gamma)^{15}$O. The $S$-factors for each transition: (b) ground state, (c) 5.18~MeV state, (d) 5.24~MeV state, (e) 6.17~MeV state, (f) 6.79~MeV state, (g) 6.86~MeV state, and (h) 7.28~MeV state. The results from previous studies~\cite{Runkle2005,imbriani2005s,bemmerer_low_2006} are also shown for comparison. The red lines are the corresponding $R$-matrix fitting results.}
\label{fig:S-factor} 
\end{figure}

Using the reaction counts extracted from BAT fitting, the $S$-factors for each transition are calculated. In the calculation, the number of $^{14}$N atoms and the beam energy loss ($\Delta E$) in the target were derived from regular yield scans of the \(E_p = 278\)~keV resonance and the trend of the target degradation, as detailed in the Supplemental Material. The $S$-factors obtained in this work were compared with data from previous studies~\cite{bemmerer_low_2006,Runkle2005,imbriani2005s} in Fig.~\ref{fig:S-factor}. The error bars represent statistical uncertainties only, which arise from the BAT fitting, beam energy calibration, $\Delta E$ determination, the non-uniformity in the target distribution, and the target degradation trends. Our total $S$-factors are consistent with the values reported by \citet{bemmerer_low_2006}, although they are systematically slightly lower. For individual transitions, the $S$-factors for the 6.17 and 6.79 MeV transitions agree with those from \citet{Runkle2005} and \citet{imbriani2005s}. The $S$-factors for the 5.18 MeV transition are consistent with \citet{imbriani2005s} except for a few low-energy points. Data for the 5.24, 6.86, and 7.28 MeV transitions are reported in this energy range for the first time. For the ground-state transition, our results are systematically higher than those of \citet{imbriani2005s} at $E_{\rm{c.m.}}<210$~keV but agree with the data from \citet{Runkle2005} at $E_{\rm{c.m.}}=180-240$~keV. 

The extrapolation of $S$-factors for each transition to zero energy was performed using an $R$-Matrix analysis with the code AZURE2~\cite{Azurma2010}. To impose basic constraints on the $R$-matrix fit, in addition to our experimental data, we included data from \citet{Runkle2005}, \citet{Li2016}, and \citet{schroder1987stellar} into the fit, thereby constraining the shapes and strengths in the higher energy regions. Due to the inconsistency at low energies, we excluded the data points from \citet{imbriani2005s} from the fit. Given that our focus is on the fitting results below the 278~keV resonance, we allowed 28 parameters that significantly affect the $S$-factors in the low-energy region to vary during the fit, based on the full set of parameters from \citet{Li2016} as the initial values. The specific parameters used in the fitting procedure are provided in the Supplemental Material. As shown in Fig.~\ref{fig:S-factor}, the $R$-matrix fits successfully reproduce both the high-energy data and the low-energy data provided by this study, with a reduced $\chi^2$ of 3.85 for all data points and 1.42 for the data points from this study. 

\begin{figure}[htpb]
\centering
\includegraphics[width=\columnwidth]{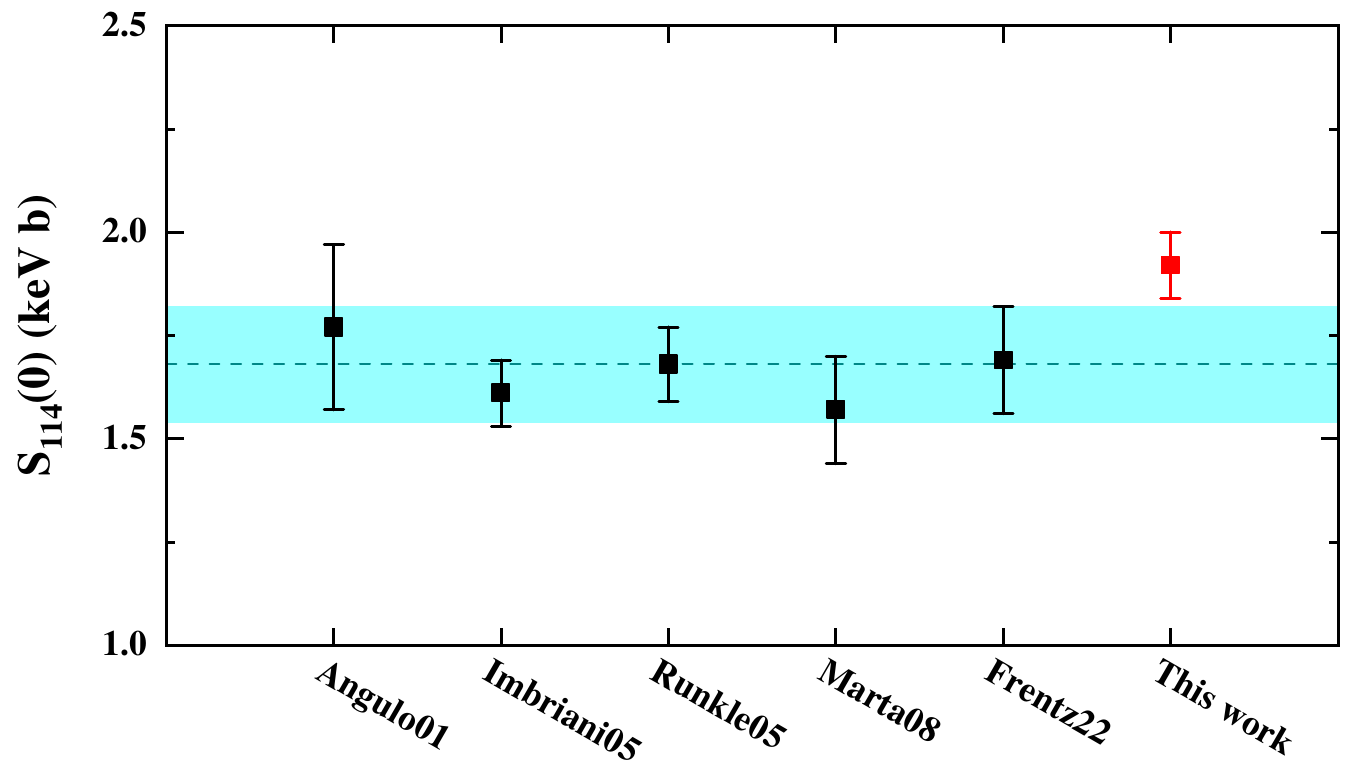}
\caption{Comparison of zero-energy $S_{114}(0)$ from different experiments~\cite{angulo200114n,imbriani2005s,Runkle2005,Marta2008,Frentz2022}. The cyan shaded area denotes the range of the value recommended in SF-III~\cite{acharya2024solar} at 1 $\sigma$ C.L.. The value of Angulo01~\cite{angulo200114n} is a $R$-matrix  analysis on the experimental data~\cite{schroder1987stellar}.}
\label{fig:S-tot} 
\end{figure}

The zero-energy total $S$-factor was determined to be $1.92\pm0.08$~keV~b, where the uncertainty includes 2.2\% statistical uncertainty and 3.4\% systematic uncertainty. This value is consistent with data reported by \citet{angulo200114n}, however, higher than those from other experiments~\cite{imbriani2005s,Runkle2005,Marta2008,Frentz2022}, as shown in Fig.~\ref{fig:S-tot}. This increase is primarily attributed to the enhancement of the ground-state transition at low energies and the inclusion of more weak transitions such as 6.86 and 7.28~MeV. Specifically, our new $S_{114}(0)$ is 14\% higher than the value recommended in SF-III~\cite{acharya2024solar}. Moreover, the improved constraints provided by the comprehensive measurement of all transitions in the $R$-Matrix fitting have significantly reduced the uncertainties in the extrapolation of the $S$-factors. The zero-energy $S$-factors for individual transitions are listed in the Supplemental Material.

To investigate the impact of the new $S_{114}$ values on the solar composition problem, the C and N abundances relative to H ($N_{\rm{CN}}$) in the photosphere of the sun were calculated following the method from previous work~\cite{Borexino2022}:
\begin{equation}
    \begin{aligned}
    \label{eq:CN}
        \frac{N_{\rm{CN}}}{N_{\rm{CN}}^{\rm{SSM}}}
        &= \frac{\left( \Phi_{\rm{O}} / \Phi_{\rm{O}}^{\rm{SSM}} \right)}
        {\left( \Phi_{\rm{B}} / \Phi_{\rm{B}}^{\rm{SSM}} \right)^{0.769}} 
        \times \left\{
        1 \pm \left[ 0.065\,(\mathrm{nucl}) 
        \right. \right. \\
        &\oplus \left. \left. 0.005\,(\mathrm{env}) \oplus 0.027\,(\mathrm{diff}) \right]
        \right\}
    \end{aligned}
\end{equation}
where $N_{\rm{CN}}^{\rm{SSM}}$ is the $N_{\rm{CN}}$ value adopted in the SSM calculation. $\Phi_{\rm{O}}$ and $\Phi_{\rm{B}}$ represent the measured fluxes of $^{15}$O and $^{8}$B neutrinos, respectively. $\Phi_{\rm{O}}^{\rm{SSM}}$ and $\Phi_{\rm{B}}^{\rm{SSM}}$ denote the corresponding values predicted by the SSM. The terms in curly brackets quantify the uncertainties due to nuclear, environmental, and diffusion effects, summed in quadrature. Compared to~\citet{Borexino2022}, the nuclear terms have been updated based on our new $S_{114}$ values and the latest $S$-factor recommendations from SF-III~\cite{acharya2024solar}.

The predicted changes in solar neutrino flux due to variations in $S_{114}$ follow a well-determined power-law relationship~\cite{villante_relevance_2021, herrera2023standard},
\begin{equation}
\begin{array}{l}
\Phi_{\mathrm{B}}^{\mathrm{new}, \mathrm{SSM}}=\Phi_{\mathrm{B}}^{\mathrm{SSM}} \times x_{S_{114}}^{0.007} \\
\Phi_{\mathrm{O}}^{\mathrm{new}, \mathrm{SSM}}=\Phi_{\mathrm{O}}^{\mathrm{SSM}} \times x_{S_{114}}^{1.051}
\end{array}
\end{equation}
where $x_{S_{114}}$ represents the ratio of the updated $S$-factor to the previously adopted $S$-factor in the SSM calculation. Therefore, the impact of our new $S_{114}$ values on the $\Phi_{\rm{B}}^{\rm{SSM}}$ and $\Phi_{\rm{O}}^{\rm{SSM}}$ can be estimated without performing additional SSM calculation. Based on the B23-GS98 predictions~\cite{herrera2023standard}, with the new $S_{114}$ values, these fluxes are updated to $\Phi_{\rm{B}}^{\rm{new, SSM}}=5.029\times10^6\ \mathrm{cm}^{-2}\mathrm{s}^{-1}$ and $\Phi_{\rm{O}}^{\rm{new, SSM}}=2.296\times10^8\ \mathrm{cm}^{-2}\mathrm{s}^{-1}$, respectively.

Combining these updated $\Phi_{\rm{B}}^{\rm{new, SSM}}$ and $\Phi_{\rm{O}}^{\rm{new, SSM}}$, the $\Phi_{\rm{B}}=5.20_{-0.10}^{+0.10}\times10^{6}~\rm{cm}^{-2}\rm{s}^{-1}$ and $\Phi_{\rm{O}}=2.53_{-0.29}^{+0.34}\times10^{8}~\rm{cm}^{-2}\rm{s}^{-1}$ from the recent global analysis of neutrino measurements~\cite{gonzalez2024status}, and the \(N_{\rm{CN}}^{\rm{SSM}}=4.143\times10^{-4}\) provided by the GS98~\cite{Grevesse1998}, we derived a new C and N abundance of \(N_{\rm{CN}}=({4.45}^{+0.69}_{-0.61})\times10^{-4}\) using Eq.~\ref{eq:CN}. This result is 23.4\% lower than the previous value \(N_{\rm{CN}}=({5.81}^{+1.22}_{-0.94})\times10^{-4}\) reported by \citet{Basilico23}, which is primarily attributed to the larger $S_{114}$ determined in this work and the lower $\Phi_{\rm{O}}$ updated in Ref.~\cite{gonzalez2024status}. Due to the improvement in the precision of the $S_{114}$ values in this work, although it remains the second contribution, the uncertainty from the nuclear cross section~(6.5\%) is now significantly smaller than that from the measured CNO neutrino flux~($^{+13.4\%}_{-11.5\%}$). 

\begin{figure}[htpb]
\centering
\includegraphics[width=\columnwidth]{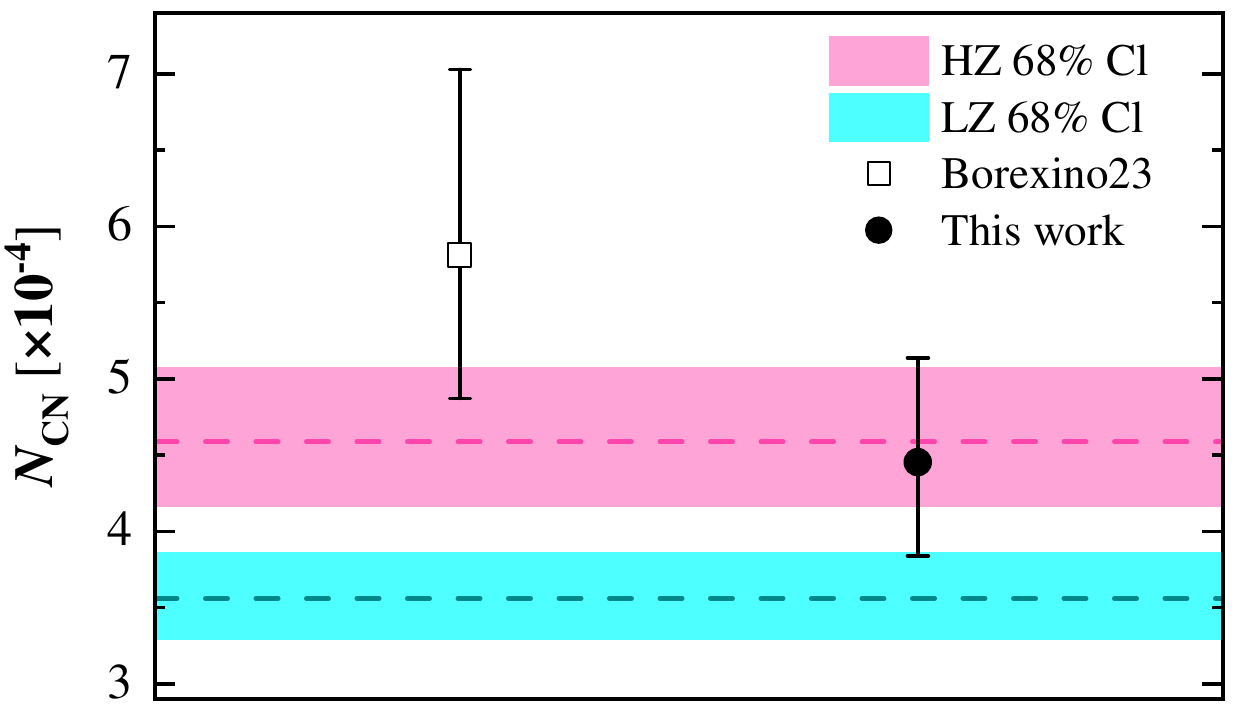}
\caption{The C and N abundance with respect to the H in the photosphere of the sun derived in the present work (solid circle). The purple and cyan shades represent the range of the HZ~\cite{magg2022observational} and LZ~\cite{asplund2021chemical} results at 1 $\sigma$ C.L., respectively. The previous result from Borexino analysis~\cite{Basilico23} (open square) is shown for comparison. }
\label{fig:NCN} 
\end{figure}

Figure \ref{fig:NCN} shows a comparison of the $N_{\mathrm{CN}}$ values derived from solar CNO neutrino measurements and those from HZ~\cite{magg2022observational} and LZ~\cite{asplund2021chemical} compositions. Compared to the previous result of \citet{Basilico23}, the updated \(N_{\rm{CN}}\) shows better agreement with the HZ value. More importantly, this new \(N_{\rm{CN}}\) value is also consistent with the LZ value within the 1$\sigma$ C.L., challenging the previous disfavor of the LZ compositions~\cite{Borexino2022,Basilico23}. Therefore, our result demonstrates that the current solar neutrino measurements are insufficient to draw a solid conclusion on the solar composition problem. Given that the uncertainty primarily arises from the measured $^{15}$O neutrino flux, further measurements of solar CNO neutrinos with higher precision are desired to address this issue. 

In summary, we conducted a new comprehensive measurement of the $^{14}$N$(p,\gamma)^{15}$O $S$-factors at low energies. Using a multiple-channel Bayesian analysis, the $S$-factors for all transitions in the energy range of $E_p=110-260$~keV were simultaneously extracted for the first time. The data for the ground-state transition, with higher precision, resolved discrepancies found in previous works. $R$-Matrix analysis of the new experimental results yields a higher and more precise $S_{114}$(0) value compared to that recommended in SF-III. The new $S_{114}$ combined with the updated B23 SSM and latest globe determination of all solar neutrino fluxes significantly reduces the $N_{\rm{CN}}$ value determined by the CNO neutrino measurements. The updated $N_{\rm{CN}}$ value agrees well with that of the HZ composition but is also consistent with the LZ composition within $1\sigma$ C.L.. As a result, we conclude that the solar composition problem remains an open question, requiring more precise observational data on solar CNO neutrino for resolution. Furthermore, given the crucial role of the $^{14}$N$(p,\gamma)^{15}$O reaction in the CNO cycle, the refined $S_{114}$ values obtained in this work are expected to have significant impacts on several related astrophysical studies.

\begin{acknowledgments}
\vspace{10pt} 
We wish to thank Institutional Center for Shared Technologies and Facilities of INEST, HFIP, CAS for operating the accelerator in Hefei. The present work is supported by the National Key R\&D Program of China (Nos.2022YFA1603301, 2022YFA1602301), National Natural Science Foundation of China (Nos.U1867211, 12275026, 12222514, 12322509, 11775133), the CNNC Science Fund for Talented Young Scholars. Y.~P.~Shen thanks R.~J.~deBoer for his support in conducting the $R$-matrix calculation using AZURE2.
\end{acknowledgments}

\bibliography{14Npg}
\end{document}


\onecolumngrid
\section*{Supplemental Material}
\subsection*{I. EXPERIMENTAL SETUP}
\begin{figure}[!ht]
\centering
\includegraphics[width=0.7\columnwidth]{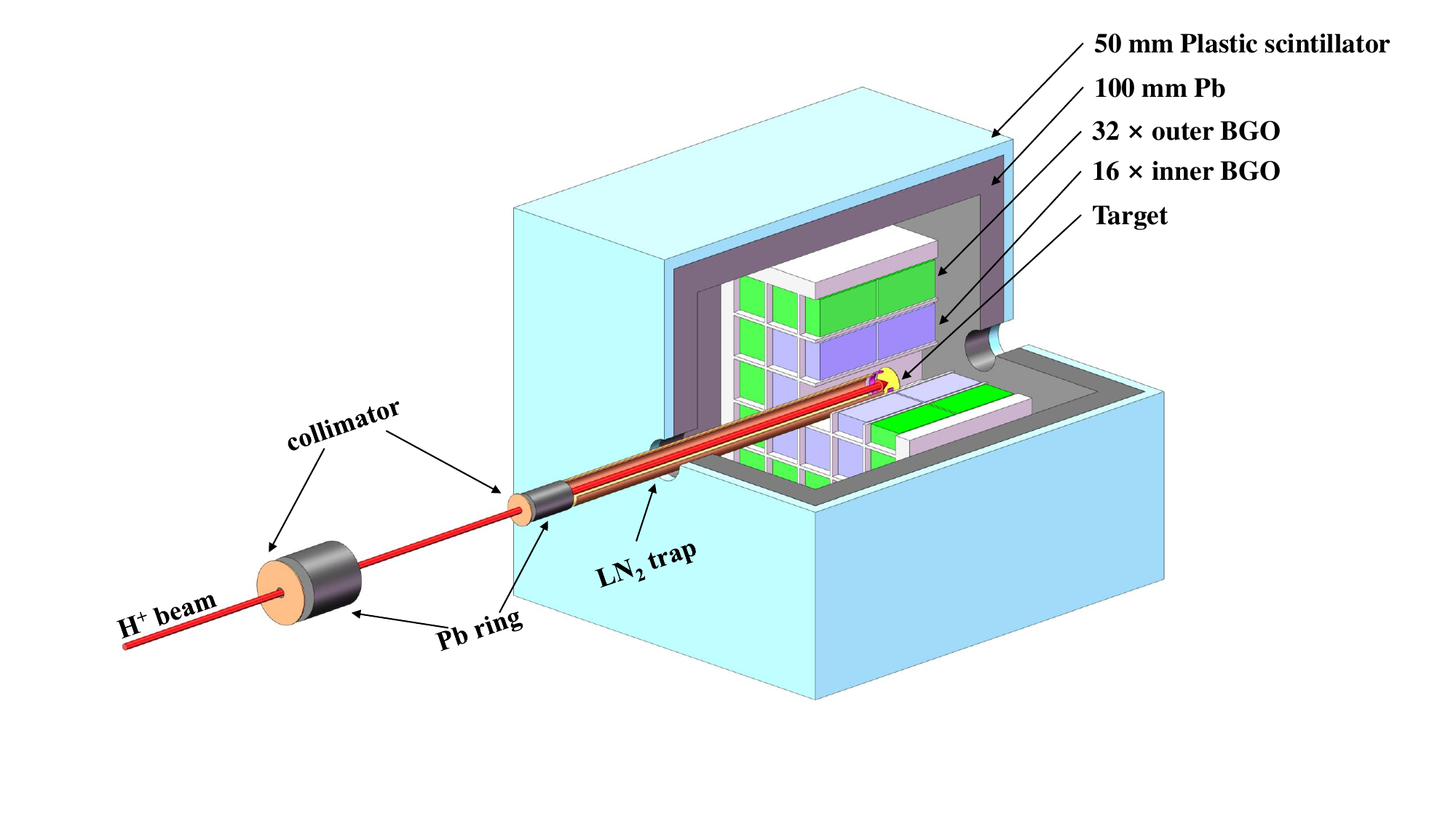}
\caption{Schematic of the experimental setup. }
\label{fig:setup} 
\end{figure}

The experiment was conducted at the 350~kV accelerator located at the Institute of Nuclear Energy Safety Technology (INEST) of the Chinese Academy of Sciences~\cite{CHEN2023168401}. This accelerator can provide a proton beam with a maximum current of 5~mA within an energy range of 50-300 ~keV. During the low-energy measurements, the typical beam current on the target was 2~mA. The high voltage of the accelerator was calibrated using the $E_p=278$~keV resonance of $^{14}$N$(p,\gamma)^{15}$O and the $E_p=151$~keV resonance of $^{18}$O$(p,\gamma)^{19}$F. Figure~\ref{fig:setup} shows the experimental setup. The H$^{1+}$ beam was collimated by two apertures ($\Phi$15~mm and $\Phi$10~mm) and directed towards TiN target with a diameter of 47.5~mm. Two Pb rings with thicknesses of 10~cm and 8~cm were installed behind the apertures to shield the $\gamma$ rays produced by the beam bombarding the apertures. The target was directly cooled using flowing deionized water to control the temperature. A liquid nitrogen-cooled cold trap was extended near the target surface to minimize the carbon deposition on the target. A ring electrode with a negative voltage of 300 V was installed at the end of the cold trap to reduce the influence of the secondary electrons on the integral beam charge. 

The second configuration of the large modular BGO detector array (LAMBDA-II)~\cite{detector} was used to measure the $\gamma$ rays emitted from the $^{14}$N$(p,\gamma)^{15}$O reaction. LAMBDA-II consists of 48 identical modules, each containing a 6~cm$\times$6~cm$\times$12~cm BGO crystal, as shown in Fig.~\ref{fig:setup} The inner 16 modules served as the primary detector array, providing nearly complete coverage of the target. The outer 32 modules acted as an anti-coincidence detector array, effectively suppressing laboratory background. We employed the summing method in measurements, where the energies of the inner 16 modules were summed to obtain the sum spectrum, and the individual spectra from each module were superimposed to form the single spectrum. Calibration using a $^{137}$Cs source resulted in an energy resolution of 10.2\% and a full energy peak efficiency of $53\pm1$\% for the 662~keV sum peak. To reduce the laboratory background, a series of shielding and anti-coincidence measures were implemented: 

1). All modules were embedded within a honeycomb-shaped block made of boron polyethylene, which effectively absorbed laboratory and cosmic ray-induced neutrons.

2). A 10~cm thick Pb shield provided full coverage to absorb the $\gamma$ ray from the outside. 

3). The five sides of the Pb shield, excluding the bottom, were covered with 5~cm thick plastic scintillators. An anti-window of 10~$\mu$s following each trigger of plastic scintillators was used to veto muons and neutrons produced by muons. 

\begin{figure}[t]
\centering
\includegraphics[width=0.7\columnwidth]{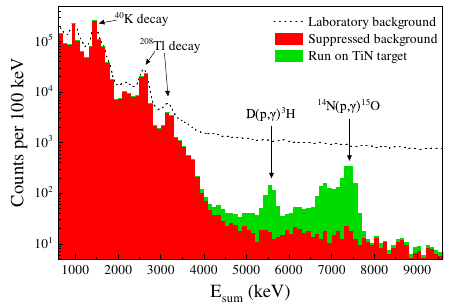}
\caption{Comparison of the sum spectrum (green shaded area) obtained at $E_p=125$~keV and the suppressed laboratory background (red shaded area). After applying the anti-coincidence method, the laboratory background under the sum peak of the $^{14}$N$(p,\gamma)^{15}$O reaction was reduced by a factor of about 60  compared to the original background level (dotted line).}
\label{fig:125 keV spectrum} 
\end{figure}

By employing the above methods, the laboratory background was reduced by about 2 orders of magnitude at $E_{\rm{\gamma}}>5$~MeV, specifically to $1\times10^{-2}$~keV$^{-1}$h$^{-1}$ around the sum peak of the $^{14}$N$(p,\gamma)^{15}$O reaction. This significant reduction in background, along with the high-intensity proton beam, enabled the extension of the $^{14}$N$(p,\gamma)^{15}$O measurement to low energies. For instance, the measured counts at the sum peak for $^{14}$N$(p,\gamma)^{15}$O at $E_{p}=125$~keV were 15 times higher than the laboratory background after applying the anti-coincidence method, as shown in Fig.~\ref{fig:125 keV spectrum}.

The beam-induced background primarily originates from the beam bombarding the impurities in the target, including $\gamma$ rays emitted by $(p,\gamma)$ reactions on D, $^{11}$B, $^{12}$C, $^{13}$C, $^{15}$N, and $^{18}$O, as well as the $(p,\alpha\gamma)$ reactions on $^{15}$N and $^{19}$F. Consequently, targets of Ti$^{15}$N, Al$_2^{18}$O$_3$, natural C, F and B were prepared via different techniques to measure the corresponding beam-induced background. The $\gamma$ rays produced by the beam bombarding the apertures are negligible due to significant attenuation by 8~cm Pb rings positioned behind them. 

\subsection*{II. TARGET PROPERTIES}
\begin{figure}[ht]
\centering
\includegraphics[width=0.7\columnwidth]{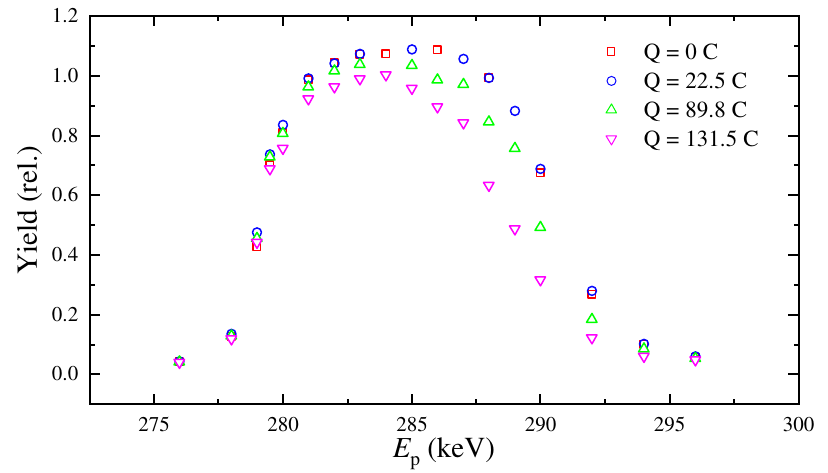}
\caption{Yield curves for the $E_p = 278$~keV resonance in the $^{14}$N$(p,\gamma)^{15}$O reaction under different amounts of charge deposited on the TiN target.}
\label{fig:scan-resonance} 
\end{figure}

To minimize impurities in the target, various techniques were employed to fabricate solid nitrogen targets, including ion sputtering, low-pressure chemical vapor deposition (LPCVD), and filtered cathodic vacuum arc (FCVA).  Among these, the TiN target prepared via FCVA exhibited the lowest impurity levels and the highest radiation resistance, making it the preferred method in this study. During the preparation, a thick Ti layer ($>1~\mu$m) was deposited on a 1~mm thick Ta substrate to prevent the proton beam from reaching the surface Ta$_2$O$_5$ layer. Subsequently, nitrogen gas was introduced into the vacuum chamber to combine with Ti ions and deposit a TiN foil on the Ti layer. The thickness of the TiN layer was adjusted by controlling the deposition time to achieve a $\sim12$~keV energy loss at $E_p$=280~keV. Details of target preparation will be discussed in the forthcoming work~\cite{target}. As shown in Fig.~\ref{fig:scan-resonance}, throughout the experiment, repeated scans of the $E_p = 278$ keV resonance yield curve for the $^{14}$N$(p,\gamma)^{15}$O reaction were conducted after accumulating a certain amount of charge on the target. This was done to monitor target degradation, nitrogen content in the TiN target, and the energy loss of the proton beam. 

The number of nitrogen nuclei in the target was calculated using the following formula:
\[
A_Y = n_s \frac{\lambda_r^2}{2} \omega \gamma
\]
where $n_s$ represents the number of active target nuclei per cm², $A_Y$ denotes the area under the resonance yield curve, $\lambda_r$ is the de Broglie wavelength corresponding to the resonance energy, and $\omega \gamma$ represents the resonance strength. In determining \(n_s\), a resonance strength of $\omega\gamma_{278} = 13.0 \pm 0.4$~meV was adopted. This value represents the weighted average of $\omega\gamma_{278} = 13.2 \pm 0.6$~meV obtained in the present study and the values reported in previous research~\cite{becker_resonance_1982, Runkle2005, imbriani2005s, bemmerer_low_2006, Daigle16, gyurky_resonance_2019, Sharma2020}, as shown in Table~\ref{table:resonance strength}. The averaging method was based on the approach used by SF~III~\cite{acharya2024solar}. The systematic uncertainty of 2.9\% associated with the nitrogen stopping power~\cite{ZIEGLER2010} was excluded from the weighted mean calculation and instead combined in quadrature with the weighted mean uncertainty to obtain the final uncertainty.


\begin{table*}[!ht]
    \centering
    \setlength{\tabcolsep}{60pt}
    \begin{minipage}{1.0\textwidth}
     \caption{Summary of resonance strength values, $\omega\gamma_{278}$, from various studies. The adopted value of $13.0 \pm 0.4$~meV represents the weighted average of the present work and previous measurements.}
    \begin{tabularx}{\textwidth}{Xc}
        \toprule
        Reference & $\omega\gamma_{278}$ (meV) \\ 
        \hline
        Becker et al. (1982)~\cite{becker_resonance_1982} & $14.1 \pm 1.0$\\ 
        Runkle et al. (2005)~\cite{Runkle2005} & $13.5 \pm 1.2$ \\ 
        Imbriani et al. (2005)~\cite{imbriani2005s} & $12.9 \pm 0.9$  \\ 
        Bemmerer et al. (2006)~\cite{bemmerer_low_2006} & $12.8 \pm 0.6$ \\ 
        Daigle et al. (2016)~\cite{Daigle16} & $12.6 \pm 0.6$ \\ 
        Gyürky et al. (2019)~\cite{gyurky_resonance_2019} & $13.4 \pm 0.8$\\ 
        Sharma et al. (2020)~\cite{Sharma2020} & $12.8 \pm 0.9$ \\ 
        \hline
        Present Work & $13.2 \pm 0.6$ \\ 
        \hline
        Adopted & $13.0 \pm 0.4$ \\ 
       \toprule
    \end{tabularx}
\label{table:resonance strength}
    \end{minipage}
\end{table*}

\subsection*{III. BRANCHING RATIOS OF THE \texorpdfstring{$E_p = 278$}{Ep = 278} KEV RESONANCE}
To accurately determine the branching ratios of resonance transitions, the angular correlation effects of $\gamma$-ray cascades between energy levels cannot be ignored. In our analysis, when using Geant4 simulations to determine the spectra of each transition component, we accounted for the theoretical angular correlation effects in the 6.17 MeV and 6.79 MeV transitions. Specifically, the angular correlation functions $W_{6.17}(\theta) = 1 + 0.138P_2(\cos \theta)$ and $W_{6.79}(\theta) = 1 + 0.250P_2(\cos \theta)$ were applied. By analyzing the results from multiple scans of the resonance for different targets, the branching ratios were finally determined, as shown in Table~\ref{tab:branch}.

\begin{table*}[ht]
    \caption{Branching ratios (in \%) of the $E_p=278$~keV resonance obtained in the present work compared with those of previous works.}
    \label{tab:branch}
    \setlength{\tabcolsep}{3mm}
    \centering
    \begin{threeparttable}
    \begin{tabular*}{\textwidth}{@{\extracolsep{\fill}}cccccc}
        \toprule
        $E_x$~(MeV) & Runkle05~\cite{Runkle2005} &Imbriani05~\cite{imbriani2005s} & Marta08~\cite{Marta2008} & Daigle16~\cite{Daigle16} & This work\\ 
        \hline
        0.00 & 1.70(7) & 1.6(1)  &  1.49(4) & 1.52(9)   & 1.65(5) \\
        5.18 & 17.3(2) & 17.1(2) & 17.3(2)  & 15.92(21) & 16.68(13) \\
        5.24 &         & 0.6(3)  & 0.15(3)  & 0.25(5)   & 0.06(6)\\
        6.17 & 58.3(5) & 57.8(3) & 58.3(4)  & 58.26(54) & 58.08(14) \\
        6.79 & 22.7(3) & 22.9(3) & 22.6(3)  & 23.86(24) & 23.39(16) \\
        6.86 &         &         &          & 0.14(4)   & 0.11(6)  \\
        7.28 &         &         &          &           & 0.05(5)   \\
        \toprule
    \end{tabular*}
     \end{threeparttable}
\end{table*}

\subsection*{\texorpdfstring{IV. PARAMETERS USED IN THE $R$-MATRIX CALCULATION}{IV. PARAMETERS USED IN THE R-MATRIX CALCULATION}}
\setlength{\LTcapwidth}{\textwidth}
\setlength{\tabcolsep}{17.8pt}
\begin{longtable}{lcccccc}
\caption{Parameters used in the R-matrix fit. The parameters varied during the fit are marked with asterisk. The signs on the partial widths and ANCs indicate the relative interferences. The dividing line demarcates the proton separation energy. The quoted uncertainties from this work are statistical only.}\\
\label{tab:r-matrixparameter}\\    
    \toprule
    \multirow{2}{*}{$E_x$~(MeV)} & \multirow{2}{*}{$J^\pi$} & \multirow{2}{*}{Channel} & \multirow{2}{*}{$l$} & \multirow{2}{*}{$s$} & \multicolumn{2}{c}{$\frac{\mathrm{ANC~fm^{-1/2}}}{\mathrm{Partial~width~(eV)}}$} \\ 
    
    \cline{6-7}
    & & & & & \citet{Li2016} & This work \\
    \hline
\endfirsthead    
    \hline
    \multirow{2}{*}{$E_x$~(MeV)} & \multirow{2}{*}{$J^\pi$} & \multirow{2}{*}{Channel} & \multirow{2}{*}{$l$} & \multirow{2}{*}{$s$} & \multicolumn{2}{c}{$\frac{\mathrm{ANC~fm^{-1/2}}}{\mathrm{Partial~width~(eV)}}$} \\
    \cline{6-7}
    & & & & & \citet{Li2016} & This work \\
    \hline
\endhead    

    \hline
    \multicolumn{7}{r}{Continued on next page} \\
    \hline
\endfoot
    \hline
\endlastfoot
    
0.0     &  $1/2^-$  &  $^{14}\mathrm{N}+\mathrm{p}$     &  1     &  1/2  & $0.23$               & $0.23$                     \\
        &           &  $^{14}\mathrm{N}+\mathrm{p}$     &  1     &  3/2  & $7.4$                & $7.4$                      \\
5.183   &  $1/2^+$  &  $^{14}\mathrm{N}+\mathrm{p}$     &  0     &  1/2  & $0.33$               & $-0.495(77)$*              \\
        &           &  $^{15}\mathrm{O}+\gamma_{0.00}$  &  $E1$  &  1/2  & $0.0784$             & $0.0784$                   \\
5.2409  &  $5/2^+$  &  $^{14}\mathrm{N}+\mathrm{p}$     &  2     &  1/2  & $0.23$               & $0.504(16)$*               \\
        &           &  $^{14}\mathrm{N}+\mathrm{p}$     &  2     &  3/2  & $0.24$               & -                          \\
        &           &  $^{15}\mathrm{O}+\gamma_{0.00}$  &  $M2$  &  1/2  & $0.0002$             & $0.0002$                   \\
6.1763  &  $3/2^-$  &  $^{14}\mathrm{N}+\mathrm{p}$     &  1     &  1/2  & $0.47$               & $0.35(2)$*                 \\
        &           &  $^{14}\mathrm{N}+\mathrm{p}$     &  1     &  3/2  & $0.53$               & $0.37(2)$*                 \\
        &           &  $^{15}\mathrm{O}+\gamma_{0.00}$  &  $M1$  &  1/2  & $0.875$              & $0.875$                    \\
6.7931  &  $3/2^+$  &  $^{14}\mathrm{N}+\mathrm{p}$     &  0     &  3/2  & $4.91$               & $4.84(2)$*                 \\
        &           &  $^{15}\mathrm{O}+\gamma_{0.00}$  &  $E1$  &  1/2  & $2.7$                & $2.873(47)$*               \\
6.8594  &  $5/2^+$  &  $^{14}\mathrm{N}+\mathrm{p}$     &  2     &  1/2  & $0.39$               & -                          \\
        &           &  $^{14}\mathrm{N}+\mathrm{p}$     &  2     &  3/2  & $0.42$               & $0.579(11)$*               \\
7.2859  &  $7/2^+$  &  $^{14}\mathrm{N}+\mathrm{p}$     &  2     &  3/2  & $1541$               & $1784(33)$*                \\
7.5565  &  $1/2^+$  &  $^{14}\mathrm{N}+\mathrm{p}$     &  0     &  1/2  & $0.96\times10^3$     & $1.19(22)\times10^3$*      \\
        &           &  $^{15}\mathrm{O}+\gamma_{0.00}$  &  $E1$  &  1/2  & $0.65\times10^{-3}$  & $0.606(66)\times10^{-3}$*  \\
        &           &  $^{15}\mathrm{O}+\gamma_{5.18}$  &  $M1$  &  1/2  & -                    & $-5.54(22)\times10^{-3}$*  \\
        &           &  $^{15}\mathrm{O}+\gamma_{5.24}$  &  $E2$  &  5/2  & -                    & $-2.9(12)\times10^{-5}$*   \\
        &           &  $^{15}\mathrm{O}+\gamma_{6.18}$  &  $E1$  &  3/2  & -                    & $2.02(42)\times10^{-2}$*   \\
        &           &  $^{15}\mathrm{O}+\gamma_{6.79}$  &  $M1$  &  3/2  & $9.3\times10^{-3}$   & $6.54(29)\times10^{-3}$*   \\
8.28514 &  $3/2^+$  &  $^{14}\mathrm{N}+\mathrm{p}$     &  2     &  1/2  & $-0.11\times10^3$    & -                          \\
        &           &  $^{14}\mathrm{N}+\mathrm{p}$     &  0     &  3/2  & $3.64\times10^3$     & $5.831(66)\times10^3$*     \\
        &           &  $^{14}\mathrm{N}+\mathrm{p}$     &  2     &  3/2  & $-0.38\times10^3$    & -                          \\
        &           &  $^{15}\mathrm{O}+\gamma_{0.00}$  &  $E1$  &  1/2  & $0.22$               & $0.2288(37)$*              \\
        &           &  $^{15}\mathrm{O}+\gamma_{5.18}$  &  $M1$  &  1/2  & -                    & $9.4(11)\times10^{-3}$*    \\
        &           &  $^{15}\mathrm{O}+\gamma_{5.24}$  &  $M1$  &  5/2  & -                    & $-0.1783(38)$*             \\
        &           &  $^{15}\mathrm{O}+\gamma_{6.18}$  &  $E1$  &  3/2  & -                    & $-1.66(14)\times10^{-2}$*  \\
8.7491  &  $1/2^+$  &  $^{14}\mathrm{N}+\mathrm{p}$     &  0     &  1/2  & $36.7\times10^3$     & $47.2(15)\times10^3$*      \\
        &           &  $^{15}\mathrm{O}+\gamma_{5.18}$  &  $M1$  &  1/2  & -                    & $0.1822(55)$*              \\
        &           &  $^{15}\mathrm{O}+\gamma_{6.18}$  &  $E1$  &  3/2  & -                    & $0.1207(41)$*              \\
8.92104 &  $5/2^+$  &  $^{14}\mathrm{N}+\mathrm{p}$     &  2     &  3/2  & $3.86\times10^3$     & $3.86\times10^3$           \\
        &           &  $^{15}\mathrm{O}+\gamma_{6.79}$  &  $M1$  &  3/2  & $3.2\times10^{-3}$   & $3.2\times10^{-3}$         \\
8.97996 &  $5/2^-$  &  $^{14}\mathrm{N}+\mathrm{p}$     &  1     &  3/2  & $-5.69\times10^3$    & $-5.69\times10^3$          \\
        &           &  $^{15}\mathrm{O}+\gamma_{0.00}$  &  $E2$  &  1/2  & $-0.29$              & $-0.29$                    \\
        &           &  $^{15}\mathrm{O}+\gamma_{6.79}$  &  $E1$  &  3/2  & $3.7\times10^{-3}$   & $3.7\times10^{-3}$         \\
9.48938 &  $5/2^-$  &  $^{14}\mathrm{N}+\mathrm{p}$     &  3     &  1/2  & $0.85\times10^3$     & $0.85\times10^3$           \\
        &           &  $^{14}\mathrm{N}+\mathrm{p}$     &  1     &  3/2  & $-7.3\times10^3$     & $-7.3\times10^3$           \\
        &           &  $^{14}\mathrm{N}+\mathrm{p}$     &  3     &  3/2  & $-1.1\times10^3$     & $-1.1\times10^3$           \\
        &           &  $^{15}\mathrm{O}+\gamma_{0.00}$  &  $E2$  &  1/2  & $-0.34$              & $-0.34$                    \\
        &           &  $^{15}\mathrm{O}+\gamma_{6.79}$  &  $E1$  &  3/2  & $0.013$              & $0.013$                    \\
9.5024  &  $3/2^+$  &  $^{14}\mathrm{N}+\mathrm{p}$     &  2     &  1/2  & $89\times10^3$       & $89\times10^3$             \\
        &           &  $^{14}\mathrm{N}+\mathrm{p}$     &  0     &  3/2  & $125\times10^3$      & $125\times10^3$            \\
        &           &  $^{15}\mathrm{O}+\gamma_{0.00}$  &  $E1$  &  1/2  & $7.4$                & $7.4$                      \\
        &           &  $^{15}\mathrm{O}+\gamma_{6.86}$  &  $M1$  &  5/2  & -                    & $0.121(12)$*               \\
9.606   &  $3/2^-$  &  $^{14}\mathrm{N}+\mathrm{p}$     &  1     &  3/2  & $-13.0\times10^3$    & $-13.0\times10^3$          \\
        &           &  $^{15}\mathrm{O}+\gamma_{0.00}$  &  $M1$  &  1/2  & $1.25$               & $1.25$                     \\
        &           &  $^{15}\mathrm{O}+\gamma_{6.79}$  &  $E1$  &  3/2  & $-0.034$             & $-0.034$                   \\
10.478  &  $3/2^+$  &  $^{14}\mathrm{N}+\mathrm{p}$     &  2     &  1/2  & $-2.4\times10^3$     & $-2.4\times10^3$           \\
        &           &  $^{14}\mathrm{N}+\mathrm{p}$     &  0     &  3/2  & $40\times10^3$       & $40\times10^3$             \\
        &           &  $^{14}\mathrm{N}+\mathrm{p}$     &  2     &  3/2  & $7\times10^3$        & $7\times10^3$              \\
        &           &  $^{15}\mathrm{O}+\gamma_{0.00}$  &  $E1$  &  1/2  & $0.38$               & $0.38$                     \\
10.497  &  $3/2^-$  &  $^{14}\mathrm{N}+\mathrm{p}$     &  1     &  1/2  & $104\times10^3$      & $104\times10^3$            \\
        &           &  $^{14}\mathrm{N}+\mathrm{p}$     &  1     &  3/2  & $-26\times10^3$      & $-26\times10^3$            \\
        &           &  $^{15}\mathrm{O}+\gamma_{0.00}$  &  $M1$  &  1/2  & $0.31$               & $0.31$                     \\
15      &  $3/2^+$  &  $^{14}\mathrm{N}+\mathrm{p}$     &  0     &  3/2  & $8.9\times10^6$      & $8.18(40)\times10^6$*      \\
        &           &  $^{15}\mathrm{O}+\gamma_{0.00}$  &  $E1$  &  1/2  & $220$                & $221(16)$*                 \\
15      &  $5/2^-$  &  $^{14}\mathrm{N}+\mathrm{p}$     &  1     &  3/2  & -                    & $2.05(18)\times10^7$*      \\
        &           &  $^{15}\mathrm{O}+\gamma_{6.86}$  &  $E1$  &  5/2  & -                    & $-118(13)$*                \\
        &           &  $^{15}\mathrm{O}+\gamma_{7.28}$  &  $E1$  &  7/2  & -                    & $74(41)$*                  \\
        \toprule
\end{longtable}


\subsection*{V. A SUMMARY OF ZERO-ENERGY S-FACTORS FOR THE $^{14}$N$(p,\gamma)^{15}$O REACTION.}
\begin{table*}[ht]
    \caption{Total zero-energy $S$-factors (keV b) and those for each transition in the $^{14}$N$(p,\gamma)^{15}$O reaction. Previous experimental results and the recommendations of SF-III are listed for comparison. The four weak transitions(5.18, 5.24, 6.86, and 7.28~MeV) are combined and shown as ``RC$\rightarrow$others''. }
    \setlength{\tabcolsep}{1mm}
    \vspace{5pt}
    \centering
    \begin{threeparttable}
    \begin{tabular*}{\textwidth}{@{\extracolsep{\fill}}cccccc}
    \toprule
        Reference & RC$\rightarrow$0 & RC$\rightarrow$6.79 & RC$\rightarrow$6.17 & RC$\rightarrow$others & Total \\ 
        \hline
        Angulo01~\cite{angulo200114n}& $0.08^{+0.13}_{-0.06}$ & $1.63\pm0.17$ & $0.06^{+0.01}_{-0.02}$   & - & $1.77\pm0.20$ \\
         Runkle05~\cite{Runkle2005} & $0.49\pm0.08$ & $1.15\pm0.05$ & $0.04\pm0.01$  & -& $1.68\pm0.09$ \\
        Imbriani05~\cite{imbriani2005s} &  $0.25\pm0.06$   &$1.21\pm0.05$  & $0.08\pm0.03$ & 0.07   &  $1.61\pm0.08$       \\
        Marta08~\cite{Marta2008} & $0.20\pm0.05$   &-  &  $0.09\pm0.07$  & -  & $1.57\pm0.13$  \\
        Frentz22~\cite{Frentz2022} & $0.33^{+0.16}_{-0.08}$        &  $1.24\pm0.09$       &  $0.12\pm0.04$       &  -         & $1.69\pm0.13$  \\
        \hline
        SF-III~\cite{acharya2024solar} & $0.30\pm0.11$        &  $1.17\pm0.03$       &  $0.13\pm0.05$       & $0.078\pm0.020$        & $1.68\pm0.14$  \\
        \hline
     This work & \makecell[c]{$0.47\pm0.04$}        &  \makecell[c]{$1.25\pm0.04$}       &  \makecell[c]{$0.11\pm0.02$}       & \makecell[c]{$0.09\pm0.02$}        & \makecell[c]{$1.92\pm0.08$}  \\
        \toprule
    \end{tabular*}
        \label{tab:br}
     \end{threeparttable}
\end{table*}


\bibliography{supp-mat}